\documentstyle[12pt,subeqn,cite]{article}
\begin{document}
\baselineskip=21pt
%
\newcommand{\bc}{\begin{center}}
\newcommand{\ec}{\end{center}}
\newcommand{\be}{\begin{equation}}
\newcommand{\ee}{\end{equation}}
\newcommand{\bq}{\begin{eqnarray}}
\newcommand{\eq}{\end{eqnarray}}
\newcommand{\bsq}{\begin{subequations}}
\newcommand{\esq}{\end{subequations}}
\begin{titlepage}
\rightline {DTP/00/81}
\rightline {gr-qc/0009054}
\bigskip
\bc
{\Large \bf  Static black holes in scalar tensor gravity}
\ec
\vskip0.5cm
\bc
{Caroline Santos\footnote{On leave from:
Departamento de F\'\i sica da Faculdade de Ci\^encias da Universidade do
Porto,
Rua do Campo Alegre 687, 4169-007 Porto, Portugal}
\vskip2cm
Centre for Particle Theory, Department of Mathematical Sciences

South Road, Durham, DH1 3LE.}
\ec

\vskip1cm
\begin{abstract}
\noindent

We study static black hole solutions in 
scalar tensor gravity. We present exact
solutions in hiperextended models with a quadratic scalar potential.

\end{abstract}

\end{titlepage}

\section{\bf Introduction}

The fact that the ratio
between the electrostatic and gravitational forces between one proton and one 
electron in the vacuum is of the same order as
the ratio between the atomic time
for the electron in its classical description
and the Hubble time, impressed Dirac in such a way that he postulated those
relations
as fundamental constants in Nature \cite{Dirac}. To keep them
independent of the cosmological time and to not 
reformulate the Atomic Physics, Dirac postulated \cite{Dirac}
a time variation for the gravitational constant, but 
his theory is in conflict with the observational results \cite{Weinberg}.
Partially motivated by the Dirac's idea and the possible existence of extra
dimensions of the space-time purposed by Kaluza and Klein \cite{Kaluza}, 
Jordan \cite{Jordan} formulated a gravity theory introducing 
the variations of the gravitational constant as an extra degree of freedom
but, again, this theory is in conflict with the observational results
\cite{Weinberg}.
These ideas culminated with the work of Brans and Dicke \cite{BDicke}
who formulated a scalar tensor theory for gravity, the so called
Jordan Brans Dicke theory.

Meanwhile Penrose and
Hawking \cite{PHawking} proved a set of theorems that
showed the existence of an initial singularity of the Universe and
of the final state of a collapsing star, named by Wheeler as black hole 
\cite{Wheeler}. 
Using a theorem due to Israel \cite{Israel} and some results from Doroshkevich,
Zel\'dovich and Novikov \cite{Wheeler}, Wheeler conjectured that 
a black hole had no hair meaning that it would
rapidly reach a stationary state uniquely determined by 
three parameters: its mass, angular
momentum and electric charge, independently of the details of the 
body that had collapsed. This conjecture was rigorously proved later
by the works of Israel, Carter, Hawking and Robinson 
\cite{ICHawkingR}.

Scalar tensor gravity seems to be the most promising
alternative to Einstein's theory of general relativity,
at least at sufficiently high energy scales.
Even this one is extremely successful at describing
the dynamics of our solar system, and indeed the 
observable universe. These theories are indistinguished
by the observational tests in the Solar system \cite{Will}
and therefore one has to look for their implications in other
regimes such as in cosmological contexts,
gravitational waves, neutron starts or black holes.
In particular Hawking studied in reference \cite{Hawking1}
static black holes in the Jordan Brans Dicke theory and comparing
them with those in Einstein gravity, he showed that they are
equivalent. This result is stated as the Hawking theorem.

In this letter we review the Hawking's theorem and 
generalize it to hiperextended Jordan Brans Dicke theories.

Therefore the layout of this letter is as follows: 
We first present some generalities for the hiperextended Jordan Brans Dicke theories.
In $Sec. III$ we review the Hawking's theorem for the 
Jordan Brans Dicke theory and in $Sec. IV$ we generalize it for some
hiperextended theories. 

\section{\bf The hiperextended Jordan Brans Dicke theories}
 
In scalar tensor theories of gravity,
the gravitational coupling is proportional to the inverse of a dynamical
scalar field \cite{BDicke},$\phi$, or in general to a function of a scalar field 
\cite{SAccetta} which
couples to the geometry by a generic coupling function, 
$\omega(\phi)$, and that can self interact in a scalar potential, $V(\phi)$.

These theories can be classified as extended or hiperextended 
depending on whether the coupling function is or is
not a constant respectively.
 
In the canonical representation, the hiperextended theories, also known as 
the generalised Jordan Brans Dicke theories, are described by the action:   
\be\label{action}
S[ g_{\mu \nu}, \phi, \chi ] = \int d^{4}x \sqrt{- g} [ \phi R 
- \frac{\omega(\phi)}{\phi} (\nabla \phi)^2
+ V(\phi) + 16 \pi L_{matter} [g_{\mu \nu}, \chi]]
\ee
where $\chi$ denotes generically nongravitational fields described by the 
Lagrangean density $L_{matter}[g_{\mu \nu}, \chi]$.

Varing the action $S$ with respect to the Jordan Brans Dicke field
and to the metric one obtains the equations of motion:
\bsq
\bq
&&R - \frac{\omega(\phi)}{\phi^{2}} (\nabla\phi)^2
+ 2 \frac{\omega(\phi)}{\phi} \Box \phi
+ \frac{dV(\phi)}{d\phi} + \frac{d\omega(\phi)}{d\phi} \frac{1}{\phi} 
(\nabla \phi)^2 = 0 \label{r} \\
&&G_{\mu \nu} = \frac{\omega(\phi)}{\phi^{2}} \left(\nabla_{\mu}\phi
\nabla_{\nu}\phi 
- \frac{1}{2}  g_{\mu \nu} (\nabla \phi)^2\right)
+ \frac{1}{\phi} \left(\nabla_{\mu}\nabla_{\nu} \phi 
- g_{\mu \nu} \Box \phi\right) 
+ \frac{V(\phi)}{2\phi} g_{\mu \nu}
+ \frac{8\pi}{\phi} T^{matter}_{\mu\nu} \label{gab}
\eq
\esq
where $T^{matter}_{\mu\nu}$ is the energy-momentum tensor
of the matter.
 
Contracting equation (\ref{gab}) and substituting the expression 
for $R$ into equation (\ref{r}) one obtains an equation for the scalar field:
\be
\Box \phi = T^{matter} \frac{8\pi}{3 + 2\omega(\phi)} 
+ \left(2 V(\phi) - \phi \frac{dV(\phi)}{d\phi}\right) 
\frac{1}{3 + 2\omega(\phi)}
- \frac{d\omega(\phi)}{d\phi} (\nabla \phi)^2
\frac{1}{3 + 2\omega(\phi)}
\ee
where $T^{matter}$ is the trace of $T^{matter}_{\mu\nu}$
and $\omega(\phi) \neq -\frac{3}{2}$ (nonsingular model).
\section{\bf The Hawking's theorem}

Let $\omega(\phi)$ be a constant and $V(\phi) = 0$ in the action
in (\ref{action}). This is the Jordan Brans Dicke action or
the action for the minimum model \cite{BDicke}.

Outside the horizon of a static black hole it is 
vacuum and therefore one gets
\be
\nabla\phi = 0\,\,\,\,.
\ee
 
Let $\varphi = \phi - \phi_{0}$ with $\phi_{0}$ the scalar 
field at far distances from the horizon. Its equation of motion is also:
\be\label{nablaphi}
\nabla\varphi = 0
\ee
 
Multiply both members of (\ref{nablaphi}) by 
$\varphi$ and integrate covariantly by parts between
two Cauchy surfaces, one placed on the horizon and the 
other on a distant region far from that. One obtains:
\be\label{integral}
\int d^{4}x \sqrt{- g} \left(\varphi_{,\alpha}\right)^{2} = 0.
\ee
 
Because $\varphi$ is a static field then $\varphi_{,\alpha}$ 
is a space time four vector and therefore
equation (\ref{integral}) implies that $\varphi_{,\alpha} = 0$,
i.e., the scalar field is constant outside the horizon
and by continuity equals to $\phi_{0}$, with \cite{Weinberg}:
\be
\phi_{0} = \frac{4 + 2\omega(\phi)}{3 + 2\omega(\phi)}\,\,\,\,.
\ee
This is the Hawking theorem \cite{Hawking1}.
\section{\bf Generalisation of the Hawking theorem}

The Hawking theorem can be immediately generalized to
extended models with $V(\phi) = V_{0}\phi^{2}$, where $V_{0}$ is a 
constant, because the equation of motion 
for the scalar field is the same as for the minimum model.

Let us now suppose  $V(\phi) = 0$, and $\omega(\phi)$ a generic 
``$well\,\,\,\, behaved$'' function. This is the Nordtvedt model
\cite{Nordtvedt}. 
Proceeding as in the previous section one concludes that
the Hawking theorem is verified when \cite{Santos}:
\be\label{constraint}
- 1 + \left(\phi - \phi_{0}\right) \frac{d\omega(\phi)}{d\phi} \frac{1}{3 
+ 2\omega(\phi)} \neq 0.
\ee
Combining these results it is immediate to generalize the
Hawking theorem to hiperextended models with $V(\phi) = V_{0}\phi^{2}$ 
and $\omega(\phi)$ a generic ``$well\,\,\,\, behaved$''function,
i.e., satisfying the constraint in (\ref{constraint}).

When the Hawking theorem is applicable as $\phi$ is constant
(equals to $\phi_{0}$) the ``$Einstein's$'' equations become:
\be
G_{\mu \nu} = \frac{V(\phi_{0})}{2\phi_{0}} g_{\mu \nu}
\ee
and the solution for the metric is given by:
\be
ds^{2} = - \left( 1 - \frac{2 M}{r} + \frac{\Lambda r^{2}}{3}\right) dt^{2} 
+ \left( 1 - \frac{2 M}{r} + \frac{\Lambda r^{2}}{3}\right)^{\small -1} dr^{2} +
r^{2} \left(d\theta^{2} + sin^{2}\theta d\psi^{2}\right)
\ee
with $\Lambda = \frac{1}{2} V_{0} \phi_{0}$. This 
is a Schwarzschild's type metric with a cosmological constant $\Lambda$.

Now let us check that this is a metric of a black hole. Calculating the 
invariant scalar $I = R_{\kappa\lambda\mu\nu}R^{\kappa\lambda\mu\nu}$, with
$R_{\kappa\lambda\mu\nu}$ the Rieman tensor curvature \cite{Wald} one obtains:
\be
I(r) = 48 \frac{M^{2}}{r^{6}} + \frac{8}{3} \Lambda^{2}
\ee
and therefore there is a singularity placed at r=0 \cite{Wald}.
The horizon is given by
\be
r_{0} = \frac{1+\sqrt[3]{(- 3 \sqrt{-\Lambda}M 
+ \sqrt{- 1 - 9 \Lambda M^{2}})^{2}}}{\sqrt{- \Lambda} 
\sqrt[3]{- 3 \sqrt{- \Lambda}M + \sqrt{- 1 - 9 \Lambda M^{2}}}}
\ee
with $M$ the mass which is greater than the critical mass $M_{critical}$:
\be
M_{critical} = \sqrt{\frac{- 1}{9 \Lambda}},
\ee
Assuming that the  Cosmic Censor Conjecture \cite{Wald} is valid one
concludes that this is indeed a black hole type singularity.
\section{\bf Acknowledgements.}

C.S. would like to thank to Mrs. Susan Percival for the reading
of this manuscript and Dr. Ruth Gregory for helpful discussions.

This work was supported by a JNICT fellowship BD/5814/95 (C.S.).

\end {document}